\documentclass[amsmath,amssymb,11pt,superscriptaddress,reprint, preprintnumbers, notitlepage,aps,twocolumn,nofootinbib]{revtex4-1}
\pdfoutput=1 
\usepackage[utf8]{inputenc}
\usepackage[english]{babel}
\usepackage{amsmath}
\usepackage{graphicx}
\usepackage{dcolumn}
\usepackage{pbox}
\usepackage{amssymb}
\usepackage{epsfig}
\usepackage[dvipsnames]{xcolor}
\usepackage[normalem]{ulem}
\usepackage{slashed}
\usepackage{amssymb}
\usepackage{ mathrsfs }
\usepackage{color}
\usepackage[font=small]{caption}
\usepackage[font=small]{subcaption}
\usepackage{url}
\usepackage[makeroom]{cancel}
\usepackage{tikz}
\usetikzlibrary{arrows}
\usepackage{enumitem} 
\usepackage{soul}

\newcommand\scalemath[2]{\scalebox{#1}{\mbox{\ensuremath{\displaystyle #2}}}}

\definecolor{MyDarkBlue}{rgb}{0.1, 0.1, 0.8} 
\definecolor{SBlue}{rgb}{0.2, 0.4, 0.7} 
\definecolor{MyLightBlue}{rgb}{0.22,0.51,0.9}
\definecolor{MyGreen}{rgb}{0.0, 0.5, 0.0}
\definecolor{BrickRed}{rgb}{0.8, 0.25, 0.33}
\RequirePackage{hyperref}
\hypersetup{colorlinks, citecolor=SBlue,linkcolor=MyDarkBlue, urlcolor=PineGreen}

\makeatletter
\renewcommand\@makecaption[2]{%
  \par
  \vskip\abovecaptionskip
  \begingroup
  
   \small\rmfamily
    \begingroup
     \samepage
     \flushing
     \let\footnote\@footnotemark@gobble
     \@make@capt@title{#1}{#2}\par
    \endgroup
  \endgroup
  \vskip\belowcaptionskip
}
\makeatother

\DeclareUnicodeCharacter{2212}{-}
\begin{document}
%%%%%%%%%%%%%%%%%%%%%%%%%%%%%%
\title{\vspace{1cm}\Large 
Is Doublet-Triplet Splitting Necessary?
}

%%%%%%%%%%%%%%%%%%%%%%%%%%%%%%
\author{\bf Ilja Dor\v{s}ner}
\email[E-mail:]{dorsner@fesb.hr}
\affiliation{University of Split, Faculty of Electrical Engineering, Mechanical Engineering and Naval Architecture in Split, Ru\dj era Bo\v{s}kovi\'{c}a 32, HR-21000 Split, Croatia}
\affiliation{J.\ Stefan Institute, Jamova 39, P.\ O.\ Box 3000, SI-1001
  Ljubljana, Slovenia}

\author{\bf Shaikh Saad}
\email[E-mail:]{shaikh.saad@unibas.ch}
\affiliation{Department of Physics, University of Basel, Klingelbergstrasse\ 82, CH-4056 Basel, Switzerland}
%%%%%%%%%%%%%%%%%%%%%%%%%%%%%%

\begin{abstract}
We demonstrate that it is not necessary to substantially break the mass degeneracy between the Standard Model Higgs boson doublet and the corresponding scalar leptoquark, where these two fields comprise a single five-dimensional $SU(5)$ representation. More precisely, we show that the experimental data on partial proton decay lifetimes cannot place any meaningful bound on the mass of the scalar leptoquark, a.k.a.\ the color triplet, if one includes effects of higher-dimensional operators. We use the Georgi-Glashow model for our demonstration and point out that the complete suppression of proton decay via the scalar leptoquark mediation is not in conflict with the partial suppression of the gauge mediated two-body proton decay signatures. Finally, we derive a proper limit on the cutoff scale of the aforementioned higher-dimensional operators and outline how this scenario could be tested. 
\end{abstract}

\maketitle
%%%%%%%%%%%%%%%%%%%%%%%%%%%%%
\section{Introduction}
%%%%%%%%%%%%%%%%%%%%%%%%%%%%%
One of the most prominent features of the 
the Georgi-Glashow model~\cite{Georgi:1974sy} is an ever-present need to introduce an enormous splitting between the masses of the two multiplets residing in the same five-dimensional scalar representation $5_H$ of the theory. One of these two multiplets is the Standard Model (SM) Higgs boson doublet $D$, whereas the other one is the color triplet $T$ that generates proton decay. Since the Higgs boson is observed to be light while the proton decay mediating scalar leptoquark should be extremely heavy, one needs to strongly break mass degeneracy between them. This issue has been referred to in the literature as the ($D$)oublet-($T$)riplet splitting problem~\cite{Dimopoulos:1981zb,Sakai:1981gr} and it persists regardless of whether the theory is supersymmetric or not.

The crux of the ($D$)oublet-($T$)riplet splitting problem in non-supersymmetric $SU(5)$ is a need to precisely cancel, i.e., fine-tune, contributions to the Higgs boson mass from $d=4$ operators $m^2 5^i_{H} 5^*_{Hi}$, $m^\prime 5^i_{H} 24_{Hi}^j 5^*_{Hj}$,  $\lambda 5^i_{H} 24_{Hi}^j 24_{Hj}^k 5^*_{Hk}$, and $\lambda^\prime 5^i_{H} 5^*_{Hi} 24_{Hk}^j 24_{Hj}^k$, where $i,j,k=1,\ldots,5$ are $SU(5)$ indices, 
$24_H$ is the scalar representation that breaks $SU(5)$ down to the SM gauge group~\cite{Georgi:1974sy}, and $m$, $m^\prime$, $\lambda$, and $\lambda^\prime$ are parameters of appropriate mass dimensions. The required cancellation takes place between contributions towards the mass of $D$ that are necessarily of the same order as the vacuum expectation value (VEV) of the SM singlet field behind the $SU(5)$ symmetry breaking in order for $T$ to remain extremely heavy.

There are several distinct approaches on how to overcome the issue of ($D$)oublet-($T$)riplet splitting within the four-dimensional $SU(5)$ framework. One approach is to implement a missing partner mechanism~\cite{Masiero:1982fe} whereby a triplet $T$ gets a large mass via coupling with an additional triplet field, whereas a doublet $D$ remains light due to an assumption that the common contributions towards masses of both $D$ and $T$ from terms such as $m^2 5^i_{H} 5^*_{Hi}$ are either all absent or very small. Note that this assumption is exactly opposite to the one that is required for the viability of the standard implementation of the ($D$)oublet-($T$)riplet splitting, but the end result is the same. There is a light multiplet $D$ and a heavy multiplet $T$, where both of these multiplets interact with the SM fermions. Since one needs to introduce additional triplet(s) without accompanying doublet(s), the missing partner mechanism implementation in $SU(5)$ calls for additional representations in the scalar sector of the theory to accomplish just that~\cite{Masiero:1982fe}.

Another possible approach to the ($D$)oublet-($T$)riplet splitting problem in $SU(5)$ is to decouple the triplet from the SM fermions, so that the triplet mass issue becomes irrelevant. Namely, triplet $T$ needs to be heavy if it yields proton decay. If one does not insist on unification of quarks and leptons of the SM, as in the original Georgi-Glashow proposal, one can couple triplet $T$ to the pairs of fermions, where one fermion is from the SM, whereas the other one is not. This can allow one to kinematically prevent proton decay in $SU(5)$, if these extra fermionic states are sufficiently heavy~\cite{Fornal:2018aqc}. This particular approach requires additional representations of both fermionic and scalar nature in order to be viable. The end result, in this case, is a triplet $T$ that can be light, but does not couple to the pairs of the SM fermions, or, at least, not to those that are relevant for preservation of matter stability. 

We will pursue an alternative direction that relies on inclusion of higher-dimensional operators and yields a triplet $T$ that exclusively couples to either the SM quark-quark or quark-lepton pairs but not to both. We will, for definiteness, use Georgi-Glashow model to provide a proof of existence of our observation although our analysis is also applicable to the $SU(5)$ theory that is broken with $75_H$~\cite{Hubsch:1984pg} instead of $24_H$, the flipped $SU(5)$ scenario~\cite{Barr:1981qv}, and supersymmetric realisations of these unification models. We will explicitly demonstrate that the inclusion of $SU(5)$ invariant higher-dimensional operators allows one to completely suppress proton decay signatures via the scalar leptoquark mediation without affecting viability of the mechanism that generates masses of the SM charged fermions. We will furthermore show that this intriguing possibility is also compatible with the partial suppression of the gauge mediated two-body proton decay signatures. 

We introduce the most relevant details of the $SU(5)$ setup and the accompanying notation in Sec.\ \ref{sec:set-up}. The proton decay suppression mechanisms for both the scalar and gauge mediations are discussed in detail in Sec.\ \ref{sec:PD}. We subsequently show how to establish an accurate upper bound on the cutoff scale associated with the higher-dimensional operators if one resorts to these suppression mechanisms in Sec.\ \ref{sec:limit}. Our findings are summarized in Sec.\ \ref{sec:conclusions}.  

%%%%%%%%%%%%%%%%%%%%%%%%%%%%%
\section{Setup}\label{sec:set-up}
%%%%%%%%%%%%%%%%%%%%%%%%%%%%%
We consider the following $d=4$ and $d=5$ operators
\begin{align}
&\mathcal{L}_Y= 
10_F^{\alpha ij}\bigg\{
Y_d^{\alpha\beta}  \overline{5}_{Fi}^\beta 5^*_{Hj} +
\frac{1}{\Lambda} Y_1^{\alpha\beta}  \overline{5}_{Fi}^\beta 5^*_{Hk} 24_{Hj}^k
\nonumber\\&+
\frac{1}{\Lambda} Y_2^{\alpha\beta}  \overline{5}_{Fk}^\beta 5^*_{Hi}   24_{Hj}^k
\bigg\}
+
10_F^{\alpha ij}10_F^{\beta kl} 5_H^m
\bigg\{
Y_u^{\alpha\beta}   \epsilon_{ijklm}
\nonumber\\&+
\frac{1}{\Lambda} Y_3^{\alpha\beta}   24_{Hm}^n \epsilon_{ijkln}
+
\frac{1}{\Lambda} Y_4^{\alpha\beta}  24_{Hk}^n \epsilon_{ijlmn}
\bigg\}+\mathrm{h.c.},
\label{eq:lagrangian}
\end{align}
where $\Lambda$ is a cutoff scale of the theory, $\alpha, \beta=1,2,3$ are family indices, $i,j,k,l,m,n=1,\ldots,5$ are $SU(5)$ indices, and $Y_d$, $Y_1$, $Y_2$, $Y_u$, $Y_3$, and $Y_4$ are Yukawa coupling matrices. Here, scalars comprise $5_H$ and $24_H$, the SM fermions reside in $\overline 5_{F}^\alpha$ and $10_{F}^\alpha$, while gauge fields are accommodated in $24_G$~\cite{Georgi:1974sy}.

The first bracket of Eq.\ \eqref{eq:lagrangian} contains operators that generate masses of the charged leptons and down-type quarks, whereas the operators from the second bracket of Eq.\ \eqref{eq:lagrangian} affect exclusively the up-type quark masses. 

The VEVs of relevance are
\begin{align}
\langle 24_H \rangle & =v_{24} \mathrm{diag}\left( -1, -1, -1, 3/2, 3/2 \right), \label{eq:24vev}
\\ \langle 5_H \rangle & = (0 \quad 0 \quad 0 \quad 0 \quad v_{5}/\sqrt{2})^T \label{eq:5vev},
\end{align}
where $v_5 \approx 246$\,GeV correctly reproduces masses of the SM gauge boson fields $W_\mu^{\pm1} \in 24_G$ and $Z^0_\mu \in 24_G$. (The field superscripts denote electric charges in units of the positron charge.)

Unification of the SM gauge couplings at scale $M_\mathrm{GUT}$, if successful, stipulates that 
\begin{align}
M_\mathrm{GUT} \equiv M_X=M_Y=\sqrt{25/8} g_\mathrm{GUT} v_{24},
\end{align}
where $g_\mathrm{GUT}$ is a gauge coupling constant at $M_\mathrm{GUT}$, while $M_X$ and $M_Y$ are masses of the proton decay mediating gauge fields $X_\mu^{\pm 4/3} \in 24_G$ and $Y_\mu^{\pm 1/3} \in 24_G$, respectively.

$\mathcal{L}_Y$ yields the following charged fermion mass matrices
\begin{align}
M_E &=v_5 \bigg\{\frac{1}{2}Y_d +\frac{3}{4} Y_1 \epsilon  - \frac{3}{4} Y_2 \epsilon \bigg\},\label{eq:uncorrected_E}\\
M_D &=v_5 \bigg\{\frac{1}{2}Y_d^T +\frac{3}{4} Y_1^T \epsilon + \frac{1}{2} Y_2^T \epsilon \bigg\}, \label{eq:uncorrected_D}\\
M_U &=v_5 \bigg\{  \sqrt{2}\left(Y_u+Y_u^T\right) + \frac{3}{\sqrt{2}}\left(Y_3+Y_3^T\right) \epsilon
 \nonumber\\&
 +  \left(\frac{1}{2\sqrt{2}} Y_4 -\sqrt{2}Y_4^T\right)\epsilon  \bigg\},\label{eq:uncorrected_U}
\end{align}
where we introduce a dimensionless parameter $\epsilon =v_{24}/\Lambda$.

The SM fermion mass matrices are diagonalized via
\begin{align}
&E^T_c M_E E=  M_E^\mathrm{diag}, \\ 
&D^T_c M_D D=  M_D^\mathrm{diag},  \\
&U^T_c M_U U=  M_U^\mathrm{diag}, \\
&N^T M_N N=  M_N^\mathrm{diag}, 
\end{align}
where $E_c$, $E$, $D_c$, $D$, $U_c$, $U$, and $N$ are $3 \times 3$ unitary matrices. $M_N$ is a $3 \times 3$ mass matrix for neutrinos, where we explicitly assume neutrinos to be Majorana particles. 

%%%%%%%%%%%%%%%%%%%%%%%%%%%%%
\section{Proton Decay Suppression}\label{sec:PD}
%%%%%%%%%%%%%%%%%%%%%%%%%%%%%
The only scalar that mediates proton decay in this scenario is $T^{-1/3}_i \equiv T_i \in 5_H^i$, where $i=1,2,3$. Its interactions with the SM fermions, as derived from Eq.\ \eqref{eq:lagrangian}, are\\
$(i)\;u^T_{k,\alpha}C^{-1}e_\beta T^*_k:$
\begin{align}
\bigg\{U^T
\bigg[
- \frac{Y_d}{\sqrt{2}}+ \frac{Y_1}{\sqrt{2}} \epsilon +\frac{3}{2\sqrt{2}} Y_2 \epsilon
\bigg] E
\bigg\}_{\alpha\beta},  \label{scalar-1}
\end{align}
\noindent $(ii)\;\epsilon_{ijk} u^{C,T}_{i,\alpha}C^{-1}d^C_{j,\beta} T^*_k:$
\begin{align}
\bigg\{U_c^\dagger
\bigg[\frac{Y_d}{\sqrt{2}}- \frac{Y_1}{\sqrt{2}} \epsilon +\frac{Y_2}{\sqrt{2}}  \epsilon
\bigg] D_c^*
\bigg\}_{\alpha\beta},  \label{scalar-2}
\end{align}
\noindent $(iii)\;d^T_{k,\alpha}C^{-1}\nu_\beta T^*_k:$
\begin{align}
\bigg\{D^T
\bigg[
\frac{Y_d}{\sqrt{2}}- \frac{Y_1}{\sqrt{2}} \epsilon -\frac{3}{2\sqrt{2}} Y_2  \epsilon
\bigg] N
\bigg\}_{\alpha\beta},  \label{scalar-3}
\end{align}
\noindent $(iv)\;\epsilon_{ijk} u^T_{i,\alpha}C^{-1}d_{j,\beta} T_k:$
\begin{align}
\scalemath{0.80}
{ 
\bigg\{U^T
\bigg[
-2\left(Y_u+Y_u^T\right) +2\left(Y_3+Y_3^T\right)\epsilon
-\frac{1}{2} \left(Y_4+Y_4^T\right)\epsilon
\bigg] D
\bigg\}_{\alpha\beta}
},  \label{scalar-4}
\end{align}
\noindent $(v)\;u^{C,T}_{k,\alpha}C^{-1}e^C_\beta T_k:$
\begin{align}
\scalemath{0.82}
{ 
\bigg\{U^\dagger_c
\bigg[
2\left(Y_u+Y_u^T\right) -2\left(Y_3+Y_3^T\right)\epsilon
+\left(3Y_4-2Y^T_4\right)\epsilon
\bigg] E_c^*
\bigg\}_{\alpha\beta}
}. \label{scalar-5}
\end{align}

Since the linear combinations of Yukawa coupling matrices that enter the charged fermion masses $M_E$, $M_D$, and $M_U$, as given in Eqs.\ \eqref{eq:uncorrected_E}, \eqref{eq:uncorrected_D}, and \eqref{eq:uncorrected_U}, respectively, differ from those that are featured in the interactions of the scalar leptoquark $T$ with the SM fermions due to the presence of $d=5$ contributions, one can suppress the latter without affecting viability of the former. For example, one can set to zero all quark-quark couplings of $T$ in Eqs.\ \eqref{scalar-2} and \eqref{scalar-4} by imposing the following three relations 
\begin{align}
  Y_2\epsilon&=Y_1\epsilon-Y_d,
\label{eq:a}\\
 Y_u+Y_u^T &= (Y_3+Y_3^T)  \epsilon \equiv Y_S,
\label{eq:b}\\
Y_4 &= -Y_4^T \equiv Y_A,
\label{eq:c}
\end{align}
where the viability of Eq.\ \eqref{eq:b} will be discussed in detail later on. These relations consequentially lead to
\begin{align}
Y_d &= \frac{4}{5 v_5} E^*_c M_E^\mathrm{diag} E^\dagger, \label{eq:Yd}\\
Y_1 &= \frac{4}{5\epsilon v_5} D^* M_D^\mathrm{diag} D^\dagger_c,\label{eq:Y1}\\
Y_S &= \frac{\sqrt{2}}{10 v_5} \left( U^*_c  M_U^\mathrm{diag} U^\dagger +U^*  M_U^\mathrm{diag} U_c^\dagger\right),
\label{eq:YS}\\
Y_A &= \frac{\sqrt{2}}{5\epsilon v_5} \left( U^*_c  M_U^\mathrm{diag} U^\dagger -U^*  M_U^\mathrm{diag} U_c^\dagger\right).\label{eq:YA}
\end{align}

Even though we still need to demonstrate that one can indeed cancel $d=4$ contributions with $d=5$ contributions in the up-type quark sector with perturbative Yukawa couplings, Eqs.\ \eqref{eq:a} through \eqref{eq:YA} already imply that generation of viable charged fermion masses is compatible with complete suppression of scalar mediated proton decay signatures. Therefore, one of the main results of this manuscript is that  experimental data on partial proton decay lifetimes cannot provide any meaningful constraint on the scalar leptoquark mass if one includes effects of higher-dimensional operators. 

If one is to include $d>5$ operators involving $5_H$ in the flavor sector of the theory, one would reintroduce quark-quark couplings with multiplet $T$. Note, however, that the viability of charged fermion masses as well as proton decay suppression can already be accommodated with $d=4$ and $d=5$ operators of Eq.\ \eqref{eq:lagrangian}. This simply means that the proton decay data can only provide constraint on the coefficients associated with $d>5$ operators as a function of scalar leptoquark mass without invalidating our result. Also, one can simply redefine interactions obtained at the $d=5$ level by incorporating the couplings generated though $d>5$ operators and proceed with aforementioned cancellations. In other words, the conditions in Eqs.\ \eqref{eq:a}, \eqref{eq:b}, and \eqref{eq:c} represent one possible flavor scenario, among many, that does not yield proton decay through the scalar exchange and thus serves as a proof of existence we are interested in. Our proposal relies on precise
cancellation, i.e., fine-tuning, as the standard implementation of the
($D$)oublet-($T$)riplet splitting but, in our case, the cancellation takes place in the flavor sector instead of the scalar sector and yields potentially light triplet $T$ that still couples to the SM fermions.

We have explicitly demonstrated how to suppress the quark-quark couplings of the scalar $T$ through cancellations of individual contributions towards its interactions with the SM fermions. We could have alternatively opted to suppress the quark-lepton couplings of Eqs.\ \eqref{scalar-1}, \eqref{scalar-3}, and \eqref{scalar-5} via the following two relations 
\begin{align}
Y_2\epsilon&=\frac{2}{3}\left(Y_d-Y_1\epsilon\right),\\
\left(Y_3+Y_3^T\right)\epsilon&=  \left(Y_u+Y_u^T\right)+\left(\frac{3}{2} Y_4-Y_4^T\right)\epsilon.
\end{align}
This is yet another way to eliminate proton decay through the color triplet mediation. Note, however, that we cannot simultaneously suppress both sets of the scalar leptoquark couplings. (This, for example, distinguishes our proposal from an approach advocated in Ref.\ \cite{Dvali:1992hc} and implemented within $SO(10)$ framework that insists on complete decoupling of the color triplet from the SM fermions, thus resulting in possible existence of long-lived colored state(s)~\cite{Dvali:1995hp}.) 

Our approach to the ($D$)oublet-($T$)riplet splitting problem, as we already stated, can be easily implemented within the supersymmetric framework. For example, within the minimal supersymmetric $SU(5)$ model, augmented with higher-dimensional operators to address observed charged fermion masses, one of the two triplets would have couplings given in Eqs.\ \eqref{scalar-1}, \eqref{scalar-2}, and \eqref{scalar-3}, whereas the other triplet would have couplings given in Eqs.\ \eqref{scalar-4} and \eqref{scalar-5}. One could then suppress either the quark-quark 
or lepton-quark interactions of both of these triplets using the conditions already discussed in our manuscript. This, in turn, would completely suppress all $d=5$ and $d=6$ proton decay operators associated with these two triplets. (There have been several attempts~\cite{Berezhiani:1998hg,Bajc:2002bv,Bajc:2002pg} within the minimal supersymmetric $SU(5)$ to exploit the freedom in the flavor sector of the theory to try to partially suppress couplings of the triplets to the SM fermions in order to alleviate tension between experimentally established proton decay lifetime limits and associated theory predictions.)

%%%%%%%%%%%%%%%%%%%%%
%%%%%%%%%%%%%%%%%%%%%

The next question we want to address is whether the complete suppression of proton decay signatures via the scalar leptoquark mediation is compatible with the partial suppression of the gauge mediated proton decay.

To that end, we recall that the relevant $d=6$ operators, which govern all eight two-body proton decay channels via exchanges of $X_\mu^{\pm 4/3}$ and $Y_\mu^{\pm 1/3}$ gauge bosons are~\cite{FileviezPerez:2004hn,Nath:2006ut}
\begin{align}
&\mathcal{O}_{I}= \frac{g_\mathrm{GUT}^2}{2M_\mathrm{GUT}^2} c(e^C_{\alpha},
d_{\beta})  \epsilon_{ijk} 
 \overline{u^C_i}  \gamma^{\mu}  u_j  \overline{e^C_{\alpha}} 
\gamma_{\mu}  d_{k \beta}, \\
&\mathcal{O}_{II}= \frac{g_\mathrm{GUT}^2}{2M_\mathrm{GUT}^2} c(e_{\alpha}, d^C_{\beta})  \epsilon_{ijk}  
\overline{u^C_i}  \gamma^{\mu}  u_j  \overline{d^C_{k \beta}} 
\gamma_{\mu}  e_{\alpha},\\
&\mathcal{O}_{III}= \frac{g_\mathrm{GUT}^2}{2M_\mathrm{GUT}^2} c(\nu_\rho, d_{\alpha}, d^C_{\beta}) 
\epsilon_{ijk}  \overline{u^C_i}  \gamma^{\mu}  d_{j \alpha}
 \overline{d^C_{k \beta}}  \gamma_{\mu}  \nu_\rho.   
\end{align}
The dimensionless coefficients of interest read
\begin{equation}
c(e^C_\alpha, d_\beta)= \left(U_c^\dagger U\right)_{11} \left(E_c^\dagger D\right)_{\alpha \beta} + ( U_c^\dagger D)_{1
\beta}( E_c^\dagger U)_{\alpha 1}, \label{gauge-1}
\end{equation}
\begin{equation}
c(e_\alpha, d_\beta^C) =   \left(U_c^\dagger U\right)_{11}\left(D_c^\dagger E\right)_{\beta \alpha},\label{gauge-2}
\end{equation}
\begin{equation}
c(\nu_\rho, d_\alpha, d^C_\beta)=  \left( U_c^\dagger D \right)_{1 \alpha}
(D_c^\dagger N)_{\beta \rho}.  \label{gauge-3}  
\end{equation}

Although proton decay via gauge boson exchange cannot be completely rotated away in $SU(5)$~\cite{Nandi:1982ew}, it is possible to significantly suppress coefficients in Eqs.\ \eqref{gauge-1}, \eqref{gauge-2}, and \eqref{gauge-3}. This, for example, can be achieved with the followings set of conditions~\cite{Dorsner:2004xa}
\begin{align}
&\left(  U^\dagger_c D  \right)_{1\alpha}=0, \label{condition-1}   
\\
&\left(  E^\dagger_c D  \right)_{1\alpha}=\left(  E^\dagger_c D  \right)_{\alpha1}=0,  \label{condition-2} 
\\
&\left(  D^\dagger_c E  \right)_{1\alpha}=\left(  D^\dagger_c E  \right)_{\alpha1}=0,  \label{condition-3}
\end{align}
where $\alpha=1,2$. Namely, since Eq.\ \eqref{condition-1} implies, via $U^\dagger D = \mathrm{diag}(e^{i \phi_1},e^{i \phi_2},e^{i \phi_3})V_\mathrm{CKM}\mathrm{diag}(e^{i \phi_4},e^{i \phi_5},1)$, that $|(U_c^\dagger U)_{11}|= |(V_\mathrm{CKM})_{13}|$, one finds that Eqs.\ \eqref{condition-1}, \eqref{condition-2}, and \eqref{condition-3} facilitate the following levels of suppression
\begin{align}
&\max|c(e^C_\alpha, d_\beta)|=2  \,\,\rightarrow\,\, \max|c(e^C_\alpha, d_\beta)|=|(V_\mathrm{CKM})_{13}|,\nonumber   
\\
&\max|c(e_\alpha, d_\beta^C)|=1 \,\,\rightarrow\,\, \max|c(e_\alpha, d_\beta^C)|=|(V_\mathrm{CKM})_{13}|,  \nonumber 
\\
&\max|c(\nu_\rho, d_\alpha, d^C_\beta)|=1 \,\,\rightarrow\,\, \max|c(\nu_\rho, d_\alpha, d^C_\beta)|=0.\nonumber
\end{align}
Here, $V_\mathrm{CKM}$ represents the Cabibbo-Kobayashi-Maskawa mixing matrix with $|(V_\mathrm{CKM})_{13}| \approx 4 \times 10^{-3}$, while $\phi_1$, $\phi_2$, $\phi_3$, $\phi_4$, and $\phi_5$ are arbitrary phases.

Even though the matrix elements of $Y_d$, $Y_1$, $Y_S$, and $Y_A$, as given in Eqs.\ \eqref{eq:Yd}, \eqref{eq:Y1}, \eqref{eq:YS}, and \eqref{eq:YA}, respectively, become somewhat more constrained after implementation of Eqs.\ \eqref{condition-1}, \eqref{condition-2}, and \eqref{condition-3}, it is still possible to completely suppress scalar mediated proton decay signatures.
Again, Eq.\ \eqref{condition-1} establishes relation between unitary matrices $U_c$ and $U$ via $V_\mathrm{CKM}$, whereas Eqs.\ \eqref{condition-2} and \eqref{condition-3} connect unitary transformations in the charged lepton and down-type quark sectors. For example, Eq.\ \eqref{condition-2} relates $E_c$ and $D$ via 
\begin{align}
&E_c = D \begin{pmatrix}
0 &0&e^{i\xi_1}\\
0&e^{i\xi_2}&0\\
e^{i\xi_3}&0&0
\end{pmatrix}, 
\end{align}
where $\xi_1$, $\xi_2$, and $\xi_3$ are arbitrary phases. Eq.\ \eqref{condition-3} introduces an analogous relationship between $D_c$ and $E$.

It is clear that simultaneous suppression of both the scalar and gauge mediation of proton decay does not only constrain Yukawa couplings of leptoquark $T$ to the SM fermions, but also introduces strong correlation between pairs of unitary matrices ($U$, $U_c$), ($D$, $E_c$), and ($E$, $D_c$).  The detection of the scalar leptoquark $T$ decay patterns at current or future colliders would consequently enable verification of consistency between predicted lepton-quark-leptoquark or quark-quark-leptoquark interactions and the observed final states.

%%%%%%%%%%%%%%%%%%%%%%%%%%%%%
\section{Cutoff Scale Limit}\label{sec:limit}
%%%%%%%%%%%%%%%%%%%%%%%%%%%%%
Let us now derive an accurate upper bound on the cutoff scale $\Lambda$ of the setup that addresses charged fermion masses via Eq.\ \eqref{eq:lagrangian} if one furthermore implements Eqs.\ \eqref{condition-1}, \eqref{condition-2}, and \eqref{condition-3} to partially suppress gauge mediated proton decay signatures. This result improves the findings of Ref.\ \cite{Dorsner:2006hw}, where the need to use $d=5$ operators to generate experimentally observed mismatch between the masses of charged leptons and down-type quarks and conditions of Eqs.\ \eqref{condition-2} and \eqref{condition-3} were exploited to find the associated upper bound on the cutoff scale. We, on the other hand, find that the most stringent limit on $\Lambda$ originates from implementation of Eq.\ \eqref{condition-1} and the fact that the leading term in the up-type quark mass matrix $M_U$ is symmetric in the flavor space, as evident from Eq.\ \eqref{eq:uncorrected_U}.

Since a symmetric form of the up-type quark mass matrix would imply $|(U_c^\dagger U)_{11}|= 1$, whereas Eq.\ \eqref{condition-1} yields $|(U_c^\dagger U)_{11}|= |(V_\mathrm{CKM})_{13}|$, one needs to have a substantial skew-symmetric component in $M_U$. To quantify this, we decompose $M_U$ into a sum of a symmetric part $S$ and a skew-symmetric part $A$ via
\begin{equation}
M_U= U^*_c M_U^\mathrm{diag} U^\dagger = \frac{v_5}{\sqrt{2}} \left(S+A\right).
\label{eq:SA}
\end{equation}
For our purposes it is sufficient to take $M_U^\mathrm{diag}=\mathrm{diag}(0,0,m_t)=\mathrm{diag}(0,0,y_t v_5/\sqrt{2})$, where $m_t$ and $y_t$ are the top quark mass and Yukawa coupling at $M_\mathrm{GUT}$ scale, respectively. (Our derivation of the limit on $\Lambda$ is accurate up to the corrections of the order of $m_c/m_t$, where $m_c$ is the charm quark mass.) This decomposition allows us to find a lower limit on the largest possible entry in $A$, i.e., $\max|(A)_{\alpha \beta}|$, in units of $y_t$. Since the elements of $A$ can only originate from the very last operator in Eq.\ \eqref{eq:uncorrected_U}, our approach should yield an accurate lower limit on parameter $\epsilon \equiv v_{24}/\Lambda$ if we insist on perturbativity of Yukawa couplings. Coincidentally, decomposition of Eq.\ \eqref{eq:SA} is compatible with the conditions in Eqs.\ \eqref{eq:b} and \eqref{eq:c}. This, in turn, will enable us to explicitly demonstrate that one can cancel $d=4$ contributions with $d=5$ contributions with perturbative couplings in the up-type quark sector. 

To perform our numerical analysis we first note that Eq.\ \eqref{condition-1} implies that
\begin{equation*}
U_c = U \mathrm{diag}(e^{i \phi_1},e^{i \phi_2},e^{i \phi_3})V_\mathrm{CKM}\mathrm{diag}(e^{i \phi_4},e^{i \phi_5},1) X^\dagger,    
\end{equation*}
where
\begin{align}
&X= \begin{pmatrix}
0 &0&e^{i\eta_4}\\
e^{i(\eta_1/2+\eta_2+\eta_3)} c_\theta & e^{i(\eta_1/2+\eta_2-\eta_3)} s_\theta &0\\
-e^{i(\eta_1/2-\eta_2+\eta_3)} s_\theta &e^{i(\eta_1/2-\eta_2-\eta_3)} c_\theta&0
\end{pmatrix}. 
\end{align}
Here, $c_\theta \equiv \cos \theta$ and $s_\theta \equiv \sin \theta$, where $\theta$ is a free parameter, while $\eta_1$, $\eta_2$, $\eta_3$, and $\eta_4$ are four arbitrary phases. (The values of elements in $V_\mathrm{CKM}$ are sourced from the Particle Data Group~\cite{ParticleDataGroup:2022pth}, although the only $V_\mathrm{CKM}$ entry that matters for our analysis is the Cabibbo angle.)
We can, furthermore, represent unitary matrix $U$ via
\begin{equation*}
U = \mathrm{diag}(e^{i \beta_1},e^{i \beta_2},e^{i \beta_3})U_{23} U_{13} U_{12} \mathrm{diag}(e^{i \beta_4},e^{i \beta_5},1),    
\end{equation*}
where 
\begin{align}
&U_{12}= \begin{pmatrix}
c_{u_{12}} & s_{u_{12}} & 0\\
-s_{u_{12}} & c_{u_{12}} & 0\\
0 & 0 &1
\end{pmatrix},\\ 
&U_{13}= \begin{pmatrix}
c_{u_{13}} & 0 & s_{u_{13}} e^{-i \beta_6}\\
0 & 1 & 0\\
-s_{u_{13}} e^{i \beta_6} & 0 & c_{u_{13}}\\
\end{pmatrix},\\ 
&U_{23}= \begin{pmatrix}
1 & 0 & 0\\
0 & c_{u_{23}} & s_{u_{23}} \\
0 & -s_{u_{23}} & c_{u_{23}} \\
\end{pmatrix}. 
\end{align}
To evaluate $U^*_c M_U^\mathrm{diag} U^\dagger$ of Eq.\ \eqref{eq:SA} and thus determine a lower limit on the largest entries in both $A$ and $S$, one needs to vary angle $\theta$ in $X$ and angles $u_{12}$, $u_{13}$, and $u_{23}$ in $U$. It also seems that one would need to vary phases $\eta_1, \ldots, \eta_4$, $\phi_1, \ldots, \phi_5$, and $\beta_1, \ldots, \beta_6$. However, it turns out that $\eta_4$ drops out due to the form of $M_U^\mathrm{diag}$, while $\eta_1$, $\eta_2$, and $\eta_3$ can be absorbed through redefinition of phases $\phi_4$ and $\phi_5$. Also, any changes in $\beta_1$, $\beta_2$, and $\beta_3$ equally affect elements $(S+A)_{\alpha \beta}$ and $(S+ A)_{\beta \alpha}$. This makes $\beta_1$, $\beta_2$, and $\beta_3$ completely irrelevant for our analysis. Finally, phases $\beta_4$ and $\beta_5$ that appear on the right-hand side of $M_U^\mathrm{diag}$ in Eq.\ \eqref{eq:SA} drop out from considerations, while $\beta_4$ and $\beta_5$ that appear on the left-hand side of $M_U^\mathrm{diag}$ can be absorbed in $\phi_1$ and $\phi_2$, respectively. After all is said and done, one only needs to vary four angles and five phases. These phases are $\phi_1$, $\phi_2$, $\phi_4$, $\phi_5$, and $\beta_6$, where $\phi_3$ is also irrelevant as it can be interpreted as an overall phase in Eq.\ \eqref{eq:SA}, upon redefinition of phases $\phi_1$ and $\phi_2$. 

\begin{figure}
\centering
\includegraphics[width=0.45\textwidth]{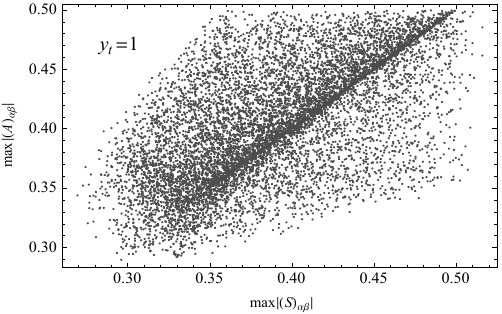}
\caption{\label{fig:SA} A plot of $\max|(A)_{\alpha \beta}|$ vs.\ $\max|(S)_{\alpha \beta}|$ for $10^4$ sets of randomly chosen values of relevant parameters. For demonstration purpose we set $y_t=1$.}
\end{figure}

We have accordingly performed a scan over $10^{6}$ configurations of the randomly chosen value sets of these nine parameters, i.e., $\theta$, $u_{12}$, $u_{13}$, $u_{23}$, $\phi_1$, $\phi_2$, $\phi_4$, $\phi_5$, and $\beta_6$, to find that $\max|(A)_{\alpha \beta}| \in (0.288 y_t,0.500 y_t)$ and $\max|(S)_{\alpha \beta}| \in (0.261 y_t,0.521 y_t)$. Moreover, $\max|(A)_{\alpha \beta}|$ and $\max|(S)_{\alpha \beta}|$ tend to be correlated. Namely, if the largest entry in $A$ is on the larger side, so is the largest entry of matrix $S$. To that end we present in Fig.\ \ref{fig:SA} a plot of $\max|(A)_{\alpha \beta}|$ vs.\ $\max|(S)_{\alpha \beta}|$ for $10^4$ randomly chosen sets of values of four angles and five phases of relevance.

To find an upper bound on $\Lambda$ we first observe that the relevant limits are $\max|(A)_{\alpha \beta}| > 0.288 y_t$ and $\max|(S)_{\alpha \beta}| > 0.261 y_t$. Also, the assumption that $Y_4$ is skew-symmetric, is consistent with the search for an upper limit on $\Lambda$ as it yields the smallest possible maximum for values of elements in $Y_4$ matrix. We can thus write, if we take $Y_4=-Y_4^T$ and use Eqs.\ \eqref{eq:uncorrected_U} and \eqref{eq:SA} to find both the symmetric and skew-symmetric contributions in $M_U$, that $S=2 (Y_u+Y_u^T) + 3 (Y_3+Y_3^T) \epsilon$ and $A\equiv 5/2 Y_4 \epsilon$.  The latter identity, then, leads to the following inequality
\begin{equation}
    \frac{5}{2} \max|(Y_4)_{\alpha \beta}| \epsilon \equiv \frac{5}{2} \max|(Y_4)_{\alpha \beta}| \frac{v_{24}}{\Lambda}> 0.288 y_t.
    \label{eq:bound_a}
\end{equation}
To proceed, we need to establish what the viable values of $v_{24}$ and $y_t$, to be used as input in Eq.\ \eqref{eq:bound_a}, are. To do that we first observe that $v_{24}=M_\mathrm{GUT} \sqrt{(2 \alpha_\mathrm{GUT}^{-1})/25 \pi}$, where $\alpha_\mathrm{GUT}=g_\mathrm{GUT}^2/(4 \pi)$. What we thus need are values of gauge coupling unification scale $M_\mathrm{GUT}$ and associated values for $\alpha_\mathrm{GUT}$ and $y_t$. 

To obtain appropriate values of $M_\mathrm{GUT}$, $\alpha_\mathrm{GUT}$, and $y_t$ we resort to a model~\cite{Dorsner:2005fq,Dorsner:2005ii,Dorsner:2006hw} that $(i)$ uses $d=4$ operator to generate neutrino masses with a single fifteen-dimensional scalar representation through the type II seesaw mechanism~\cite{Lazarides:1980nt,Mohapatra:1980yp}, $(ii)$ relies on operators of Eq.\ \eqref{eq:lagrangian} to generate charged fermion masses, and $(iii)$ implements Eqs.\ \eqref{condition-1}, \eqref{condition-2}, and \eqref{condition-3} to suppress gauge mediated proton decay signatures. This model yields  $M_\mathrm{GUT}=4.5 \times 10^{14}$\,GeV, $\alpha_\mathrm{GUT}^{-1}=37.1$ and $y_t=0.42$, where $\alpha_\mathrm{GUT}$ is evaluated via the two-loop level gauge coupling unification analysis, whereas $y_t$, at $M_\mathrm{GUT}$, is obtained through the one-loop level renormalisation group equation running. If these values for $\alpha_\mathrm{GUT}$ and $y_t$ are inserted into Eq.\ \eqref{eq:bound_a} we find that the upper limit on $\Lambda$ for a model of Ref.\ \cite{Dorsner:2005fq} reads $\Lambda < 57 M_\mathrm{GUT}$, where we assume that $5/2 \max|(Y_4)_{\alpha \beta}| = \sqrt{4 \pi}$.  This bound on $\Lambda$, as inferred from Eq.\ \eqref{eq:bound_a}, is another key finding of this manuscript. It is, as already advocated, much more stringent than the $\Lambda \lesssim 900 M_\mathrm{GUT}$ bound reported in Ref.\ \cite{Dorsner:2006hw}. 

We note that it has been recently argued~\cite{Senjanovic:2024uzn} that the particle content of the Georgi-Glashow model~\cite{Georgi:1974sy} is already sufficient to address gauge coupling unification if one adds $d=5$ operators in the gauge sector of the theory~\cite{Shafi:1983gz}. That proposal~\cite{Senjanovic:2024uzn} also relies on suppression of gauge mediated proton decay signatures for its viability but uses $d=5$ operator to accommodate neutrino masses. We find that the most stringent cutoff scale limit of that model actually originates from the need to have viable mass scale for neutrinos. Consequentially, the use of Eq.~\eqref{eq:bound_a} to find associated cutoff within that model would not be appropriate. 

It is now trivial to show that it is possible to cancel $d=4$ contributions with $d=5$ terms in the up-type quark sector, as required by Eq.\ \eqref{eq:b}, if one is to completely suppress scalar mediated proton decay signatures. Namely, if we insert conditions of Eqs.\ \eqref{eq:b} and \eqref{eq:c} into Eq.\ \eqref{eq:uncorrected_U} and implement numerical analysis of the decomposition of Eq.\ \eqref{eq:SA}, we find that
\begin{equation}
    5 \max|(Y_3+Y_3^T)_{\alpha \beta}| \epsilon > 0.261 y_t.
    \label{eq:bound_b}
\end{equation}
Since the right-hand side of Eq.\ \eqref{eq:bound_a} is larger than the one in Eq.\ \eqref{eq:bound_b}, we see that one can simultaneously accomplish complete suppression of the scalar mediated proton decay and partial suppression of the gauge mediated proton decay, where the correct upper bound on the cutoff scale $\Lambda$ in Eq.\ \eqref{eq:bound_a} originates from Eqs.\ \eqref{eq:uncorrected_U} and \eqref{condition-1}.

Let us, as a qualitative example, assume that $\epsilon=0.05$. This, for $\alpha_\mathrm{GUT}^{-1}=37.1$ and $y_t=0.42$, implies that $\Lambda=19.4 M_\mathrm{GUT}=8.7 \times 10^{15}$\,GeV, $5 \max|(Y_3+Y_3^T)_{\alpha \beta}| > 2.2$, $5 \max|(Y_4)_{\alpha \beta}|/2 > 2.4$, and $2 \max|(Y_u+Y_u^T)_{\alpha \beta}| > 0.88$ if one is to implement suppression of both scalar and gauge mediations of proton decay within a model of Refs.\ \cite{Dorsner:2005fq,Dorsner:2005ii,Dorsner:2006hw}.

If one only wants to suppress scalar mediated proton decay, one can completely drop a skew-symmetric contribution towards $M_U$ in Eq.\ \eqref{eq:uncorrected_U}. This could be accomplished by setting, for example, $Y_4=0$. In this scenario, $M_U=M_U^T$ implies $U_c \equiv U$. We can thus resort to the decomposition of Eq.\ \eqref{eq:SA}, drop $A$, set $U_c \equiv U$, and vary three angles in $U$, together with three phases $\beta_4$, $\beta_5$, and $\beta_6$, to establish correct upper limit on the cutoff scale $\Lambda$. This procedure, combined with the condition of Eq.\ \eqref{eq:b}, yields  
\begin{equation}
    5 \max|(Y_3+Y_3^T)_{\alpha \beta}| \epsilon > 0.334 y_t.
    \label{eq:bound_c}
\end{equation}
It is clear that if one drops a skew-symmetric contribution to $M_U$, one would obtain somewhat more stringent limit on the cutoff scale $\Lambda$. It is thus preferable to have the $d=5$ skew-symmetric contribution towards $M_U$ or even obligatory if one is to suppress proton decay signatures through  gauge boson mediation. Amazingly, though, it is entirely possible to completely suppress proton decay through scalar mediation even when the up-type quark mass matrix is symmetric as long as $\Lambda$ satisfies Eq.\ \eqref{eq:bound_c}.

%%%%%%%%%%%%%%%%%%%%%%%%%%%%%
\section{Conclusions}\label{sec:conclusions}
%%%%%%%%%%%%%%%%%%%%%%%%%%%%%
We demonstrate that without additional assumptions about the flavor structure of the theory, the experimental data on partial proton decay lifetimes cannot meaningfully constrain the mass of the scalar leptoquark in $SU(5)$ if one includes effects of higher-dimensional operators. Our result is applicable to both supersymmetric and non-supersymmetric $SU(5)$ frameworks that resort to the use of $d=5$ operators to provide viable masses for the SM charged fermions. This undermines the underlying premise behind the need to implement the doublet-triplet mass splitting in this class of models.

We also show that the complete suppression of proton decay via the scalar leptoquark mediation is compatible with the partial suppression of the gauge mediated proton decay signatures.

Finally, we show how to derive an accurate upper bound on the cutoff scale of the $d=5$ operators in the fermion sector if one implements partial suppression of the two-body proton decay signatures due to the gauge boson exchanges, complete suppression of the scalar mediated proton decay, or both. 

%%%%%%%%%%%%%%%%%%%%%%%%%%%%%
\section*{Acknowledgments}
%%%%%%%%%%%%%%%%%%%%%%%%%%%%%
I.D.\ acknowledges the financial support from the Slovenian Research Agency (research core funding No.\ P1-0035).

%%%%%%%%%%%%%%%%%%%%%%%%%%%
%%%%%%%%%%%%%%%%%%%%%%%%%%%
\bibliographystyle{style}
\bibliography{reference}
\end{document}